\documentclass[twocolumn,showpacs,preprintnumbers,amsmath,amssymb]{revtex4}

\usepackage{graphicx}
\usepackage{dcolumn}
\usepackage{bm}
\usepackage{textcomp}
\usepackage{wasysym}
\usepackage{stmaryrd}

\def \ee{\end{equation}}
\def \be{\begin{equation}}

\preprint{}

\begin{document}

\title{Relativistic Bohmian mechanics from scalar gravity}

\author{Benjamin Koch}
 \affiliation{
 Pontificia Universidad Cat\'{o}lica de Chile, \\
Av. Vicu\~{n}a Mackenna 4860, \\
Santiago, Chile \\
}

\date{\today}

\begin{abstract}
In this article we show that the
fundamental equations of relativistic Bohmian mechanics
for a single particle
can be derived from a scalar theory of curved space-time.
This paper is written in honor of
David Bohm.
\end{abstract}

\maketitle

%
\section{Introduction}

The de Broglie-Bohm interpretation (dBB) offers an alternative way for
the intuitive understanding of quantum mechanical phenomena in
terms of particle trajectories \cite{Bohm:1951,Bohm:1951b}.
At the same time it gives a clear answer to the typical
difficulties that appear in the standard interpretation of quantum mechanics, 
like the so called collapse of the wave function.
This theory was further generalized to relativistic quantum mechanics
and quantum field theory with bosonic and fermionic fields
\cite{Bell:1988,Holland:1985ud,Nikolic:2002mi,Struyve:2006cj}.
The above approaches address the critical issues
that are involved with the dBB and its relativistic generalization,
such as non-locality, non-conservation of particle numbers and
the theory of quantum measurements. 
However, they all start by postulating a set of three differential
equations which include the famous ``quantum potential''.
In this paper we show that those three equations of relativistic bosonic
quantum mechanics naturally follow from a scalar theory of
curved space-time similar to that suggested by 
Gunnar Nordstr{\"o}m \cite{Nordstrom:1913a,Nordstrom:1913b,Ravndal:2004ym}.
Since this result is valid for a consistent interpretation of
the quantum Klein Gordon equation, it is also valid in the non-relativistic
limit, giving the Schr{\"o}dinger equation.
This work is based on a previous attempt on
a similar subject \cite{Koch:2008hn}, but the previous inconsistencies
and technical problems have now been overcome.
The idea of combining geometry and the quantum Klein-Gordon
equation has already been suggested in a number of
different approaches:\\
An analogous formulation of our geometrical toy model
can also be obtained by studying a nontrivial vector transplantation law
in Weyl geometry \cite{Santamato:1984qe}. 
However, this was done without giving
the necessary relation to Bohmian mechanics.\\
A different approach 
studies conformal transformations which
directly contain the dBB quantum potential $Q$
\cite{Shojai:2000us,Carroll:2004hs}.
Further, the Einstein-Hilbert
action is extended by kinetic terms for two scalar fields
and a constraint condition which identifies one
of those fields with the quantum potential is imposed.
The highly coupled differential equations of
this approach contain 
general relativity as a zero order approximation
in a parameter  $\sim \hbar$.
The single particle dBB equations
result from a first order approximation.
Thus, in contrast to the toy model which is presented
here, the Bohmian equations are not part of an
exact duality.\\
Recently a study on the Ricci flow equation has independently shown that
the Ricci scalar can be identified with a
potentail \cite{Carroll:2007zh,Abraham:2008yr}
(in our case the Schr{\"o}dinger quantum potential).
This is consistent with the first condition in (\ref{eq_match}).
%
\section{Particle trajectories in relativistic Bohmian mechanics}

In this section we shortly list the ingredients for
the interpretation of the single particle quantum Klein-Gordon equation
in terms of Bohmian trajectories. 
For a detailed description of subsequent topics in the dBB 
theory like particle creation,
the theory of quantum measurement,
many particle states, and quantum field theory
the reader is referred to \cite{Nikolic:2006az}.
The first two equations postulated in this
version of the dBB theory are
\begin{eqnarray}\label{eq_KG1}
0&\equiv&
\partial_m \left(P^2 (\partial^m S_Q) \right)\quad, \\ 
\label{eq_KG2}
2 M Q&\equiv&(\partial^m S_Q)(\partial_m S_Q)
-M^2\quad,
\end{eqnarray}
where  $M$ is the particle mass, $S_Q$ is the quantum phase,
$Q$ is  the quantum potential, and $P=P(x)$ is the pilot wave.
The first equation takes the form of a continuity equation
(which gives in the non-relativistic limit probability conservation)
and the second equation has the form of a classical Hamilton-Jacobi equation
with the additional potential $Q$.
The use of the Latin indices in contrast
to the usual Greek indices is in order to
allow a distinction between coordinates in
curved space-time (Greek) and coordinates in flat space-time (Latin).
The quantum potential is further given from the particle mass and
the pilot wave by
\be\label{eq_QuantPot}
Q=\frac{\hbar^2}{2M} \frac{\partial^m \partial_m P}{P}\quad.
\ee
Note that the two real equations (\ref{eq_KG1},\ref{eq_KG2})
can be rearranged by using the equation (\ref{eq_QuantPot})
and the definition $\Phi(x,t)\equiv P \exp(i S_Q/\hbar)$ in such
a way that they give the complex quantum Klein-Gordon equation
\be\label{eq_KG0}
 \partial^m \partial_m\Phi(x)=-\frac{M^2}{\hbar^2}\Phi(x) \quad.
\ee
The interpretation in terms of particle trajectories \cite{Nikolic:2006az}
(which differs from the interpretation given by \cite{Holland:1985ud}) 
enters now
by the third postulate, which relates the particle momentum $p^m$
to the quantum phase
\be\label{eq_KG3}
p^m = M\frac{d x^m}{d\tau}\equiv\partial^m S_Q \quad.
\ee
A derivative of this equation with the identity
\be\label{eq_der1}
\frac{d}{ds}=\frac{dx^m}{ds}\partial_m \quad,
\ee
gives the equation of motion of a relativistic particle 
in the dBB interpretation
\be\label{eq_eom}
\frac{d^2x^m}{ds^2}=\frac{(\partial^nS_Q)(\partial^m\partial_n S_Q)}{M^2}\quad.
\ee
By using equation (\ref{eq_KG2}) a further simplification of this equation
can be made 
\be
M\frac{d^2x^m}{ds^2}=\partial^m Q \quad.
\ee
Note that the described theory gives in
the non-relativistic limit the original dBB interpretation
of the Schr{\"o}dinger equation \cite{Nikolic:2002mi}.
%

\section{Scalar theory of curved space-time}

Now we will discuss 
a scalar theory of curved space-time, similar to the theory of gravitation
suggested by Gunnar Nordstr{\"o}m \cite{Nordstrom:1913a,Nordstrom:1913b}. 
This theory does not predict light bending due
to heavy astronomical objects and therefore it is not in agreement with 
astrophysical observations.
Although it is now known
that this theory can not reproduce gravity it yields
some very interesting features which we will develop in this section.
The two defining equations are
\be\label{eq_Nord1}
R=\kappa T\quad,
\ee
and
\be\label{eq_Nord2}
C_{\mu\nu\alpha\beta}=0\quad,
\ee
where $R$ is the Ricci scalar, $T$ is the trace of the stress-energy
tensor, 
$\kappa$ is the coupling constant of this theory, 
and 
$C_{\mu\nu\alpha\beta}=R_{\mu\nu\alpha\beta}-(g_{\mu[\alpha}R_{\beta]\nu}
-g_{\nu[\alpha}R_{\beta]\mu})+1/3Rg_{\mu[\alpha}g_{\beta]\nu}$ 
is the Weyl curvature tensor. 
One further imposes that the metric has a vanishing covariant derivative
\be\label{eq_metcon}
(\nabla_\kappa g)_{\mu \nu}=0\quad.
\ee
We will now show that it is possible
to derive the equations of the dBB theory 
(\ref{eq_KG1}- \ref{eq_KG3}, and \ref{eq_eom})
from equations (\ref{eq_Nord1}, \ref{eq_Nord2}, and \ref{eq_metcon})
by demanding one set of matching conditions.
From the conditions (\ref{eq_Nord2}, \ref{eq_metcon}) one can
show that the connection $\Gamma$ is of the Levi Civita type
\be\label{eq_Levi}
\Gamma^\alpha_{\mu \nu}=\frac{1}{2}g^{\alpha \beta}\left(
\partial_\mu g_{\beta \nu}+\partial_\nu g_{\beta \mu}-
\partial_\beta g_{\mu \nu}\right) \quad.
\ee
In general relativity the Levi Civita connection is derived
by imposing (\ref{eq_metcon}) and vanishing torsion.
Another consequence of the conditions (\ref{eq_Nord2}, \ref{eq_metcon})
is that the metric has to be proportional to the Minkowski metric
\be\label{eq_gmn}
g_{\mu \nu}(x)=\phi^2(x) \eta_{m n}\quad,
\ee
where the function $\phi$ is a local scale factor.
This ensures that all possible solutions for the metric are conformal
to the Minkowski space-time.
The inverse of this metric is
$g^{\mu \nu}(x)=1/(\phi^2(x)) \eta^{m n}$.
Like in the previous section the Latin and Greek indices help to
distinguish between the effective Minkowski space-time (with $\eta_{mn}$)
and the actual space-time (with $g_{\mu\nu}$).
As a consequence derivatives in both notations can be identified
$\partial_\mu=\partial_m$ but adjoint derivatives carry an additional factor
of $1/\phi^2$ since
\be\label{eq_gmn2}
\partial^\mu=g^{\mu \nu} \partial_\nu=1/\phi^2 \eta^{m n}\partial_n=
1/\phi^2\partial^m\quad.
\ee
For the stress-energy tensor $T_{\mu\nu}$
we extend the Hamilton Jacobi definition
by adding the product of a constant four vector $k_\mu$
\be\label{eq_Tmn}
T_{\mu\nu}\equiv  k_\mu k_\nu+(\partial_\mu S_H)(\partial_\nu S_H)\quad,
\ee
where the Hamilton principle function $S_H$ defines the local momentum
$p_\mu=M \,dx^\mu/d\tau=\partial_\mu S_H$.
The additional term is part of the definition of this
theory and it will in the end allow to define
matching conditions between $k_\mu$ and $M$.
Note that elevating $S_H$ to a field, corresponds
to taking into account the congruence
of possible solutions of the equations of motion
instead of just considering 
one single solution which parts from 
one single initial condition.
Two energy units ($\delta^3(x)/m$) were absorbed in this definition, 
thus the tensor carries in natural units $[GeV^2]$ and 
therefore $\kappa$ is dimensionless.
Contracting equation (\ref{eq_Tmn}) with the inverse
of equation (\ref{eq_gmn}) gives
\be
T=\frac{k^m k_m}{\phi^2(x)}+ \frac{(\partial^m S_H)(\partial_m S_H)}{\phi^2(x)}\quad.
\ee
Using the above in equation (\ref{eq_Nord1})
one finds
\be\label{eq_Nord1b}
\frac{\partial^m\partial_m \phi(x)}{\phi(x)}=-
\frac{\kappa}{6}(\partial^m S_H)(\partial_m S_H)-
\kappa \frac{ k^m k_m}{6}\quad.
\ee
Now one can see that this is exactly equation (\ref{eq_KG2})
if one identifies
\begin{eqnarray}\label{eq_match}
\phi(x)&=&P(x)\quad, \\
S_H(x)&=& S_Q(x)\;\;, \nonumber \\
\kappa&=&-6/\hbar^2\;\;, \nonumber \\
k^m k_m&=&-M^2\quad. \nonumber
\end{eqnarray}
Since the derivative of the Hamilton principle function defines
a momentum, the second condition ($S_H=S_Q$)
ensures that the third
postulate of relativistic Bohmian mechanics (\ref{eq_KG3})
is fulfilled naturally.
It is of course possible to shift the above
matching conditions into each other without
disturbing the overall matching between (\ref{eq_KG2})
and (\ref{eq_Nord1b}).
The first equation of the dBB theory (\ref{eq_KG1}) now 
follows from the condition that
the four momentum $p^\mu$ (which appears in the stress-energy tensor) 
has to be a conserved quantity
\be\label{eq_cond2}
\nabla_\mu p^\mu=\partial_\mu p^\mu +\Gamma^\nu_{\mu \nu} p^\mu=0 \quad.
\ee
Writing again the momentum in terms of the Hamilton principle function $S_H$ 
and by using (\ref{eq_Levi}, \ref{eq_gmn}),
this condition reads
\be\label{eq_cond3}
\frac{1}{\phi^4(x)}\partial_m
\left( \phi^2 \partial^m S_H\right)=0\quad.
\ee
By the virtue of the matching conditions (\ref{eq_match})
this is exactly the first Bohmian equation (\ref{eq_KG1}).
(Please note that further demanding $\nabla_\mu T^{\mu \nu}$ contains condition 
(\ref{eq_cond3}) but it gives also an additional condition:
$0= P^m((\partial_m P^n)\phi^2-1/2 P_m (\partial^n \phi^2))$.)
The total derivative (\ref{eq_der1}) is now generalized to
\be\label{eq_der2}
\frac{d}{d\tau}=\frac{dx^\mu}{d\tau}\partial_\mu=
\frac{1}{\phi^2} \frac{dx^m}{d\tau}\partial_m=
\frac{1}{\phi^2}\frac{d}{ds}\quad.
\ee
Applying this to the momentum $p^\mu$ gives the
equation of motion in curved space-time
\be\label{eq_eom2}
\frac{d^2 x^\mu}{d\tau^2}=\frac{(\partial^n\partial_n S_H)(\partial^mS_H)+
(\partial^m\partial_n S_H)(\partial^nS_H)}{M^2 \phi^4}\quad,
\ee
or equivalently the equation of motion in the Minkowski coordinates $x^m$
\be\label{eq_eom22}
\frac{d^2 x^m}{ds^2}=\frac{(\partial^m\partial_n S_H)(\partial^nS_H)}{M^2}\quad.
\ee
Comparing (\ref{eq_eom22}) to (\ref{eq_eom}) shows that
also the equation of motion of relativistic Bohmian mechanics is reproduced
from the matching conditions (\ref{eq_match}). However,
in addition to the equation of motion (\ref{eq_eom2}) there
is also the geodesic equation of motion in curved space-time
\be\label{eq_eom3}
\frac{d^2 x^\mu}{d\tau^2} +
 \Gamma^\mu_{\alpha \beta} \frac{dx^\alpha}{d\tau}\frac{dx^\beta}{d\tau}=
 f(x) \frac{dx^\mu}{d\tau}\quad,
\ee
where $f(x)$ can be some arbitrary scalar function.
Now one has to show that both equations
of motion (\ref{eq_eom2}, \ref{eq_eom3}) are consistent. 
This can be proven by plugging
equation (\ref{eq_eom2}) into (\ref{eq_eom3}) 
and by using equations (\ref{eq_Levi}, \ref{eq_eom2}), and 
by using twice
the rule
\be
\partial_m\phi^2 \partial^nS_H=-\phi^2 (\partial_m\partial^nS_H)+Y^n_m\quad,
\ee
which is equivalent to (\ref{eq_cond3}) al long as the
arbitrary matrix $Y^n_m$ is traceless.
After some transformations we find that 
\be
f(x)=-\frac{3}{2}\frac{ (\partial_a\phi^2) (\partial^aS_H)}{M^3 \phi^4}\quad,
\ee
which is indeed a scalar function.
Thus, we have shown that the straight forward equation of motion (\ref{eq_eom2})
is consistent with the geodesic equation of motion (\ref{eq_eom3}).

Before going to the conclusions we want to comment on
the matching conditions between the relativistic dBB theory and
the scalar theory of curved space-time.
Those conditions make it possible to show that
relativistic dBB theory for a single particle 
is dual to the above theory of scalar gravity.
Although the matching conditions we have chosen in (\ref{eq_match})
are not unique, they have a number of intrinsic properties
which have to be true for any set of matching conditions:
First, the scalar function $\phi$ in front of the metric (\ref{eq_gmn})
has to be proportional to the pilot wave $P$ in the dBB theory.
Second, the Hamilton principle function $S_H$ in the stress-energy
tensor has to be proportional to the quantum phase $S_Q$ of the dBB theory.
Third, if the quantum phase $S_Q$ and the Hamilton principle function $S_H$
are defined as real functions, then the coupling of the geometrical theory has to
be negative.
\section{Summary}

In this paper we showed
that the equations of the relativistic dBB theory
for a single particle
(\ref{eq_KG1}-\ref{eq_KG3}, and \ref{eq_eom})
can be derived from a scalar theory of curved space-time based on the four
equations (\ref{eq_Nord1}-\ref{eq_metcon}, and \ref{eq_Tmn}).
By ``derived'' we mean that for every function and every constant
in the dBB theory we could define a corresponding function and a corresponding
constant in the latter theory which obeys the same equations and therefore is suitable
for describing the same physical reality.
The important question whether this duality can be
generalized to the many particle case, interacting theories,
fermionic degrees of freedom, or quantum field theory will be subject of future studies.\\ \\
The author wants to thank Jorge Noronha, Marco Aurelio Diaz, Davi Rodrigues, Maximo Ba{\~n}ados,
and Hrvoje Nikolic for their comments and remarks.


\end{document}